\input{epsf}
\documentclass[11pt]{article}
\def\be{\begin{equation}}
\def\eea{\end{eqnarray}}
\def\bea{\begin{eqnarray}}
\def\ee{\end{equation}}

\author{M.Mohammadi$^{}$ \footnote{majid471702@yahoo.com}
\\$^{}$ {\small Department of Physics, Islamic Azad University - Shahreza Branch, Shahreza, Isfahan, Iran}
}
\title{Non-Markovian Analysis of the Phase Damped  Jaynes-Cummings
           Model in the Presence of a Classical Homogeneous Gravitational Field
 }
 \begin{document}
\maketitle
\begin{abstract}
\noindent In this paper, the non-Markovian dissipative dynamics of the phase damped Jaynes-Cummings
           model in the presence of a classical
           homogeneous gravitational field will be analyzed. The model consists of a moving two-level atom simultaneously
exposed to the gravitational field and a single-mode traveling radiation field in the presence of a
non-Markovian phase damping mechanism.
 First, the non-Markovian master equation for the reduced density operator of the system in terms of a
 Hamiltonian describing the atom-field interaction in the presence of a homogeneous gravitational field will be presented.
 Then, the super-operator technique will be generalized and an exact solution of the non-Markovian master equation will be
  obtained.
 Assuming that initially the radiation
field is prepared in a Glauber coherent state and the two-level atom is
in the excited state, the non-Markovian effects
 on the temporal evolution of collapses and revivals of the atomic
population inversion and photon counting statistics of
the radiation field in the presence of both the phase damping and a homogeneous gravitational field will be investigated.
\end{abstract}
\\
\section{Introduction}
Non-Markovian effects have received special attention in the past years, mainly in optics and radiation-matter interaction
subjects, either in predicting novel effects or due to the necessity to go beyond the Markovian approximation in experiments
involving femtosecond processes. Among the experimental papers the recent ones can be cited. Tchenio et al. prepared a
Non-Markovian atomic excitation process, with adjustable memory time, using correlated laser pulses and they verified that
under strong-field conditions the atoms are not able to keep memory of the field phase and amplitude over a time interval
larger than the coherence time [1]. Considering femtosecond experiments, non-Markovian behavior appears in the optical
dephasing of molecules in solution, since the dynamics of the thermalized environment may occur on the same time
scale of the system [2-6]. Concerning the theoretical approach, Lewenstein et al. predicted the suppression
of spontaneous emission related to the decay of cavity atoms in the presence of a strong driving field, thus modifying
the spectrum of resonance fluorescence [7]. Villaeys et al. studied the non-Markovian effects in the atomic
absorption band shape for the transient and steady-state regimes; they conclude that in the steady state the
 appearance of the non-Markovian effects are washed out and therefore they cannot be probed, but in the
  transient regime these effects are perceptible [8]. In this same line Gangopadhyay and Pay constructed
  a non-Markovian master equation by considering density matrices with small delay time [9].
  The prototype of such systems, proposed by the Jaynes-Cummings model (JCM) in 1963, [10] describes a two-level
atom resonantly interacting with a single-mode quantized field. It has proved to be a theoretical laboratory of great
relevance to many topics in atomic physics and quantum optics, as well as in the ion traps, cavity QED and quantum
information processing [11]. When the rotating wave approximation (RWA) is made, the model becomes exactly
solvable and its dynamical
features can be analytically brought to light revealing remarkable properties [12].
In the standard JCM, the interaction between a
constant electric field and a stationary (motionless) two-level
atom is considered. With the development in the technologies of
laser cooling and atom trapping the interaction between a moving
atom and the field has attracted much attention [13-22].\\
\hspace*{00.5 cm} Experimentally, atomic beams
with very low velocities are generated in laser cooling and
atomic interferometry [23]. It is obvious that for atoms moving
with a velocity of a few millimeters or centimeters per second
for a time period of several milliseconds or more, the influence
of Earth's acceleration becomes important and cannot be neglected
[24]. A semi-classical description of a two-level atom interacting with a
running laser wave in a gravitational field has been studied
[25,26]. However, the semi-classical treatment does not permit us
to study the pure quantum effects occurring in the course of
atom-radiation interaction. Recently, within a quantum
treatment of the internal and external dynamics of the atom, we have presented [27] a theoretical scheme based on
an su(2) dynamical algebraic structure to investigate the
influence of a classical homogeneous gravitational field on the quantum
non-demolition measurement of atomic momentum in the dispersive
JCM. Also, the effects of the
gravitational field on quantum statistical properties of the
lossless [28] as well as the phase-damped JCMs [29] were investigated. We reach to the point
that the gravitational field seriously suppresses non-classical
properties
of both the cavity-field and the moving atom. Also, the effects of the
gravitational field on the dynamical evolution of the cavity-field
entropy and the creation of the Schr\"{o}dinger-cat state in the
Jaynes-Cummings model [30] are examined.\\
\hspace*{00.5 cm} On the other hand, over the last two decades much attention has been focused
on the properties of the dissipative variants of the JCM. The theoretical
efforts have been stimulated by experimental progress in cavity QED. Besides the experimental drive,
there also exists a theoretical motivation to include relevant damping mechanism
to JCM because its dynamics becomes more interesting. A number of authors have treated the JCM with
 dissipation by the use of analytic approximations [31,32] and numerical calculations [33-37]. The
solution in the presence of dissipation is not only of theoretical interest, but also important from a practical
point of view since dissipation would be always present in any experimental realization of the model.
 However, the dissipation treated in the above studies is modeled by coupling to an external reservoir including
 energy dissipation. As is well known, in a dissipative quantum system, the system loses energy by creating
 a bath quantum. In this kind of damping the interaction Hamiltonian between bath and system does not commute with the
 system Hamiltonian. In general this leads to a thermalization of the system with a certain time constant.
 There are, however other kinds of environmental coupling to the system, which do not involve energy exchange.
 In the so-called phase damping [38] the interaction Hamiltonian commutes with that of system and
 in the dynamics only the phase of system state is changed in the course of interaction. Similar to standard energy damping the
 off-diagonal elements of the density matrix in energy basis decay at a given rate. The phase damping can well describe
 some unaccounted decay of coherences in a single-mode micromaser [39]. It has also been shown that phase damping
 seriously reduces the fidelity of the received qubit in quantum computers due to the induced decoherence
 [40]. The phase damping in the JCM with one quantized field mode has been studied [41]. The influence of phase damping
 on non-classical properties of the multi-quanta two-mode JCM has also been studied [42]. It has been found
 that the phase damping suppresses non-classical effects of the cavity field in the JCM. However, all of the foregoing
 studies have been done only under the condition that the influence of the gravitational field is not taken into account.\\
 \hspace*{00.5 cm} In this paper, the non-Markovian dissipative dynamics of the phase damped Jaynes-Cummings
           model in the presence of a classical
           homogeneous gravitational field will be analyzed. The model consists of a moving two-level atom simultaneously
exposed to the gravitational field and a single-mode traveling radiation field in the presence of a
non-Markovian phase damping mechanism. In sect.2, the non-Markovian master equation
for the reduced density operator
of the system in terms of a Hamiltonian describing the atom-radiation interaction in
the presence of a gravitational field will be presented. This Hamiltonian has been obtained based on an
 su(2) dynamical algebraic structure in the interaction picture. In sect.3 an exact solution of
  the JCM with the phase damping in the presence of a gravitational field will be obtained, by which
  the dynamical evolution of the system is investigated.
 In sect.4 the non-Markovian effects on both the cavity-field and the atomic properties
 in the presence of both the phase damping and a homogeneous gravitational field will be studied.
Considering the field to be initially in a coherent state and the
two-level atom in the excited state, the temporal evolution of the
atomic inversion and photon
counting statistics will be explored. Finally, our conclusions will be summarized in section 5.\\
\section{ Non-Markovian Master Equation for the Phase Ddamped JCM in the Presence of a Gravitational Field }
The equation of motion for the density operator of the atom-radiation
 system and reservoir, $\hat{\rho}_{sr}(t)$,
  in the Schr\"{o}dinger picture is given by [29]
\begin{equation}
\frac{\partial\hat{\rho}_{sr}(t)}{\partial t}=-i[\hat{\tilde{H}}_{T},\hat{\rho}_{sr}(t)]  (\hbar=1),
\end{equation}
where
\begin{equation}
\hat{\tilde{H}}_{T}=\hat{H}_{s}+\hat{H}_{r}+\hat{V}_{sr},
\end{equation}
with the Hamiltonian of the reservoir
\begin{equation}
\hat{H}_{r}=\sum_{i}\omega_{i}\hat{b}_{i}^{\dag}\hat{b}_{i},
\end{equation}
and with the Hamiltonian of the interaction between the system and reservoir
\begin{equation}
\hat{V}_{sr}=\hat{H}_{s}\sum_{j=1}^{3}\hat{F}_{j},
\end{equation}
where
\begin{equation}
\hat{F}_{1}=\sum_{i}\kappa_{i}\hat{b}_{i}, \hat{F}_{2}=\sum_{i}\kappa_{i}\hat{b}_{i}^{\dag}, \hat{F}_{3}=\hat{H}_{s}\sum_{i}\frac{\kappa_{i}^{2}}{2\omega_{i}},
\end{equation}
$\hat{b}_{i}$ and $\hat{b}_{i}^{\dag}$ are the boson annihilation and creation operators for the reservoir
 and $\kappa_{i}$ is
the coupling constant.
The Hamiltonian $\hat{H}_{s}$ in (2) for the atom-radiation system in the presence of a classical gravity
 field with the atomic
motion along the position
vector $\hat{\vec{x}}$ and in the RWA is given by ($\hbar=1$)
\begin{eqnarray}
\hat{H}_{s}=&&\frac{\hat{p}^{2}}{2M}-M\vec{g}.\hat{\vec{x}}+\omega_{c}(\hat{a}^{\dag}\hat{a}+\frac{1}{2})+\frac{1}{2}\omega_{eg}\hat{\sigma}_{z}+\nonumber\\
&&\lambda[\exp(-i\vec{q}.\hat{\vec{x}})\hat{a}^{\dag}\hat{\sigma}_{-}+\exp(i\vec{q}.\hat{\vec{x}})\hat{\sigma}_{+}\hat{a}],
\end{eqnarray}
where $\hat{a}$ and $\hat{a}^{\dag}$ denote, respectively, the
annihilation and creation operators of a single-mode traveling
wave with frequency $\omega_{c}$, $\vec{q}$ is the wave vector of
the running wave and $\hat{\sigma}_{\pm}$ denote the raising and
lowering operators of the two-level atom with electronic levels
$|e\rangle, |g\rangle $ and Bohr transition frequency
$\omega_{eg}$. The atom-field coupling is given by the parameter
$\lambda$ and
 $\hat{\vec{p}}$, $\hat{\vec{x}}$
denote, respectively, the momentum and position operators of the
atomic center of mass motion and $g$ is Earth's gravitational
acceleration. It has been shown [29] that based on an su(2)
algebraic structure, as the dynamical symmetry group of the
model, in the interaction picture, Hamiltonian (6) takes the following form
\begin{eqnarray}
\hat{\tilde{H}}^{I} _{s}=&&\omega_{c}(\hat{a}^{\dagger}\hat{a}+\frac{\hat{S}_{0}}{2})+\frac{1}{2}\hat{\Delta}(\hat{\vec{p}},\vec{g},t)\hat{S}_{0}\\ \nonumber +&& (\hat{\kappa}(t)
\sqrt{\hat{K}}\hat{S}_{-}+\hat{\kappa}^{*}(t)\sqrt{\hat{K}}\hat{S}_{+}),
\end{eqnarray}
where
the operators
\begin{equation}
\hat{S_{0}}=\frac{1}{2}(|e \rangle \langle e|-|g \rangle \langle
g|) , \hat{S_{+}}=\hat{a}|e\rangle \langle
g|\frac{1}{\sqrt{\hat{K}}},
\hat{S_{-}}=\frac{1}{\sqrt{\hat{K}}}|g\rangle \langle
e|\hat{a}^{\dag},
\end{equation}
with the following commutation relations
\begin{equation}
[\hat{S_{0}},\hat{S_{\pm}}]=\pm
\hat{S_{\pm}},[\hat{S_{-}},\hat{S_{+}}]=-2\hat{S_{0}},
\end{equation}
are the generators of the su(2) algebra, the operator $\hat{K}=\hat{a}^{\dag}\hat{a}+|e\rangle\langle e| $
is a constant of motion which represents the total number of excitations of the atom-radiation,
  $\hat{\kappa}(t)$ is an effective coupling coefficient
\begin{equation}
\hat{\kappa}(t)= \lambda
\exp(\frac{it}{2}(\hat{\triangle}(\hat{\vec{p},t},\vec{g})+\frac{\hbar
q^{2} }{M})),
\end{equation}
 and the operator
\begin{equation}
\hat{\triangle}(\hat{\vec{p},t},\vec{g})=\omega_{c}-(\omega_{eg}+\frac{\vec{q}.\hat{\vec{p}}}{M}+\vec{q}.\vec{g}t+\frac{
q^{2}}{2M}),
\end{equation}
has been introduced as the Doppler shift detuning at time $t$
[29]. By using following the same procedure as in refs.[29,44] we obtain the non-Markovian master equation
 for the reduced density operator of the system
 with neglecting $2\pi i \frac{d}{d\omega}[J(\omega)|\kappa(\omega)|^{2}]|_{\omega=0}$ and the lamb shift term
\begin{eqnarray}
\frac{\partial\hat{\rho}_{s}(t)}{\partial t}=&&-i[\hat{\tilde{H}}^{I} _{s},\hat{\rho}_{s}(t)]-\gamma [\hat{H}^{I} _{s},[\hat{H}^{I} _{s},\hat{\rho}_{s}(t)]] \\ \nonumber - && \eta [\hat{H}^{I} _{s},[\hat{H}^{I} _{s},[\hat{H}^{I} _{s},[\hat{H}^{I} _{s},\hat{\rho}_{s}(t)]]]],
\end{eqnarray}
where $\hat{\tilde{H}}^{I} _{s}$ is given by (7). In Eq.(12), $\gamma$ and $\eta$ are the damping and the non-Markovian
parameters, respectively, which depend on the temperature $T$
\begin{equation}
\gamma=2\pi T lim_{\omega\rightarrow 0}(\frac{J(\omega)|\kappa(\omega)|^{2} }{\omega}),
\end{equation}
and
\begin{equation}
\eta=2\pi T \wp\int_{0}^{\infty}d\omega J(\omega) \frac{|\kappa(\omega)|^{2}}{\omega^{3}},
\end{equation}
where $J(\omega)$ and $\kappa(\omega)$ are the spectral density
 of the reservoir and the coupling coefficient, respectively and $\wp$ is
 the Cauchy principal part of the integration [44].\\
\section{Dynamical Evolution of the Non-Markovian Phase Damped JCM in the Presence of Classical Gravity}
In section 2, we reached to the non-Markovian master equation
for the reduced density operator of the atom-radiation system in the presence of a classical homogeneous gravitational
field. In this section,
we now start to find the exact solution for the density operator $\hat{\rho}_{s}(t)$ of the non-Markovian
master equation (12)
with the Hamiltonian (7). For this purpose, the approach presented in refs.[43-45] is applied.
The formal solution is given by
\begin{equation}
\hat{\rho}_{s}(t)=\exp(\hat{R}t)\exp(\hat{S}t)\exp(\hat{T}t)\hat{\rho}_{s}(0),
\end{equation}
\begin{equation}
\exp(\hat{R}t)=\exp(\hat{R_{1}}t)\exp(\hat{R_{2}}t)\exp(\hat{R_{3}}t),
\end{equation}
\begin{equation}
\exp(\hat{T}t)=\exp(\hat{T_{1}}t)\exp(\hat{T_{2}}t),
\end{equation}
where $\hat{\rho}_{s}(0)$ is the density operator of the initial atom-field system. The auxiliary
super-operators $\hat{R_{1}}$,$\hat{R_{2}}$,$\hat{R_{3}}$, $\hat{S}$ and $\hat{T_{1}}$,$\hat{T_{1}}$
 are defined through their action on the density operator such that
\begin{equation}
\exp(\hat{R_{1}}t)\hat{\rho}_{s}(0)\equiv \sum_{k=0}^{\infty}\frac{(2\gamma t)^{k}}{k!}(\hat{\tilde{H}}^{I}_{s})^{k}\hat{\rho}_{s}(0)(\hat{\tilde{H}}^{I}_{s})^{k},
\end{equation}
\begin{equation}
\exp(\hat{R_{2}}t)\hat{\rho}_{s}(0)\equiv \sum_{l=0}^{\infty}\frac{(-3\gamma \eta t)^{l}}{l!}(\hat{\tilde{H}}^{I}_{s})^{2l}\hat{\rho}_{s}(0)(\hat{\tilde{H}}^{I}_{s})^{2l},
\end{equation}
\begin{equation}
\exp(\hat{R_{3}}t)\hat{\rho}_{s}(0)\equiv \sum_{m=0}^{\infty}\frac{(-\gamma \eta t)^{m}}{m!}(\hat{\tilde{H}}^{I}_{s})^{m}\hat{\rho}_{s}(0)(\hat{\tilde{H}}^{I}_{s})^{2m},
\end{equation}
\begin{equation}
\exp(\hat{S}t)\hat{\rho}_{s}(0)\equiv \exp(-i\hat{\tilde{H}}^{I}_{s}t)\hat{\rho}_{s}(0)\exp(i\hat{\tilde{H}}^{I}_{s}t),
\end{equation}
\begin{equation}
\exp(\hat{T_{1}}t)\hat{\rho}_{s}(0)\equiv \exp(-\gamma (\hat{\tilde{H}}^{I}_{s})^{2} t)\hat{\rho}_{s}(0)\exp(-\gamma (\hat{\tilde{H}}^{I}_{s})^{2} t).
\end{equation}
\begin{equation}
\exp(\hat{T_{2}}t)\hat{\rho}_{s}(0)\equiv \exp(-\gamma \eta(\hat{\tilde{H}}^{I}_{s})^{4} t)\hat{\rho}_{s}(0)\exp(-\gamma \eta(\hat{\tilde{H}}^{I}_{s})^{4} t).
\end{equation}
\hspace*{00.5 cm}It is assumed that initially the radiation field is in a coherent
superposition of the Fock states, the atom is in the excited state $|e\rangle$, and the state
vector for the center-of-mass degree of freedom is
$|\psi_{c.m}(0)\rangle=\int d^{3}p \phi(\vec{p})|\vec{p}\rangle$.
Therefore, the initial density operator of the atom-radiation system reads as
\begin{equation}
\hat{\rho}_{s}(0)=\hat{\rho}_{field}(0)\otimes \hat{\rho}_{atom}(0)\otimes\hat{\rho}_{c.m}(0)= \left [\begin{array}{cccc}
\hat{\rho}_{field}(0)\otimes\hat{\rho}_{c.m}(0) & 0 \\
0 & 0 \\
\end{array}\right],
\end{equation}
where
\begin{equation}
\hat{\rho}_{field}(0)=\sum_{n}\sum_{m}w_{n}(0)w_{m}(0)|n\rangle \langle m |,
\end{equation}
\begin{equation}
\hat{\rho}_{c.m}(0)=\int d^{3}p \int d^{3}p'\phi^{*}(\vec{p'})\phi(\vec{p})|\vec{p}\rangle \langle\vec{p'} |,
\end{equation}
with $w_{n}(0)=\frac{\exp(-\frac{|\alpha|^{2}}{2})\alpha^{n}}{\sqrt{n!}}$.
The Hamiltonian (7) can be expressed as a sum of two terms which commute with each other, that is,
\begin{equation}
\hat{\tilde{H}}^{I}_{s}=\hat{H}_{1}+\hat{H}_{2},[\hat{H}_{1},\hat{H}_{2}]=0
\end{equation}
where
\begin{equation}
\hat{H}_{1}=\omega_{c}(\hat{a}^{\dagger}\hat{a}+\frac{\hat{S}_{0}}{2}),
\end{equation}
\begin{equation}
\hat{H}_{2}=\frac{1}{2}\hat{\Delta}(\hat{\vec{p}},\vec{g},t)\hat{S}_{0}+ (\hat{\kappa}(t)
\sqrt{\hat{K}}\hat{S}_{-}+\hat{\kappa}^{*}(t)\sqrt{\hat{K}}\hat{S}_{+}).
\end{equation}
In the two-dimensional atomic basis we have
\begin{equation}
\hat{H}_{1}=\omega_{c} \left [\begin{array}{cccc}
\hat{n}+\frac{1}{2} & 0 \\
0 & \hat{n}-\frac{1}{2} \\
\end{array}\right],
\end{equation}
\begin{equation}
\hat{H}_{2}= \left [\begin{array}{cccc}
\frac{\Delta(\vec{p},\vec{g},t)}{4} & \kappa^{*}(t)\hat{a}  \\
\kappa(t) \hat{a}^{\dagger} & -\frac{\Delta(\vec{p},\vec{g},t)}{4} \\
\end{array}\right].
\end{equation}
Also, the square of the Hamiltonian (7) can be expressed as a sum of two operators, one of them is diagonal, in the form
\begin{equation}
(\hat{\tilde{H}}^{I}_{s})^{2}=\hat{A}_{1}+\hat{A}_{2},
\end{equation}
where
\begin{eqnarray}
\hat{A}_{1}=&&\hat{H}_{1}^{2}+\hat{H}_{2}^{2} \\ \nonumber  =&&\left [\begin{array}{cccc}
\omega_{c}^{2}(\hat{n}+\frac{1}{2})^{2}+\lambda^{2}(\hat{n}+1)+(\frac{\Delta(\vec{p},\vec{g},t)}{4})^{2} & 0 \\
0 & \omega_{c}^{2}(\hat{n}-\frac{1}{2})^{2}+\lambda^{2}\hat{n}+(\frac{\Delta(\vec{p},\vec{g},t)}{4})^{2} \\
\end{array}\right],
\end{eqnarray}
and
\begin{equation}
\hat{A}_{2}=2\hat{H}_{1}\hat{H}_{2}=2\omega_{c} \left [\begin{array}{cccc}
(\hat{n}+\frac{1}{2})(\frac{\Delta(\vec{p},\vec{g},t)}{4}) & (\hat{n}+\frac{1}{2})\kappa^{*}(t)\hat{a} \\
(\hat{n}-\frac{1}{2})\kappa(t) \hat{a}^{\dagger} & -(\hat{n}-\frac{1}{2})(\frac{\Delta(\vec{p},\vec{g},t)}{4}) \\
\end{array}\right].
\end{equation}
It is easily proved that $[\hat{A}_{1},\hat{A}_{2}]=0$.
Similarly, the square of the $(\hat{\tilde{H}}^{I}_{s})^{2}$ can be expressed as a sum of two operators,
 one of them is diagonal, in the form
\begin{equation}
(\hat{\tilde{H}}^{I}_{s})^{4}=\hat{A}_{3}+\hat{A}_{4},
\end{equation}
where
\begin{equation}
\hat{A}_{3}=\hat{A}_{1}^{2}+\hat{A}_{2}^{2}   =\left [\begin{array}{cccc}
(\hat{A}_{3})_{11}& 0 \\
0 &(\hat{A}_{3})_{22} \\
\end{array}\right],
\end{equation}
with
\begin{eqnarray}
(A_{3})_{11}=&&[\omega_{c}^{2}(\hat{n}+\frac{1}{2})^{2}+\lambda^{2}(\hat{n}+1)+(\frac{\Delta(\vec{p},\vec{g},t)}{4})^{2}]^{2}\\ \nonumber
+&&4\omega_{c}^{2}(\hat{n}+\frac{1}{2})(\frac{\Delta(\vec{p},\vec{g},t)}{4})+4\omega_{c}^{2}\lambda^{2}(\hat{n}+\frac{1}{2})^{2}(\hat{n}+1),
\end{eqnarray}
\begin{eqnarray}
(A_{3})_{22}=&&[\omega_{c}^{2}(\hat{n}-\frac{1}{2})^{2}+\lambda^{2}\hat{n}+(\frac{\Delta(\vec{p},\vec{g},t)}{4})^{2}]^{2}\\ \nonumber
+&&4\omega_{c}^{2}(\hat{n}+\frac{1}{2})(\frac{\Delta(\vec{p},\vec{g},t)}{4})+4\omega_{c}^{2}\lambda^{2}(\hat{n}-\frac{1}{2})^{2}\hat{n},
\end{eqnarray}
and
\begin{equation}
\hat{A}_{4}=2\hat{A}_{1}\hat{A}_{2}= \left [\begin{array}{cccc}
 (\hat{A}_{4})_{11}&(\hat{A}_{4})_{12}  \\
 (\hat{A}_{4})_{21}&(\hat{A}_{4})_{22}  \\
\end{array}\right].
\end{equation}
with
\begin{equation}
(\hat{A}_{4})_{11}=[\omega_{c}^{2}(\hat{n}+\frac{1}{2})^{2}+\lambda^{2}(\hat{n}+1)+(\frac{\Delta(\vec{p},\vec{g},t)}{4})^{2}]4\omega_{c}(\hat{n}+\frac{1}{2})(\frac{\Delta(\vec{p},\vec{g},t)}{4}),
\end{equation}
\begin{equation}
(\hat{A}_{4})_{12}=[\omega_{c}^{2}(\hat{n}+\frac{1}{2})^{2}+\lambda^{2}(\hat{n}+1)+(\frac{\Delta(\vec{p},\vec{g},t)}{4})^{2}]4\omega_{c}(\hat{n}+\frac{1}{2})\kappa^{*}\hat{a},
\end{equation}
\begin{equation}
(\hat{A}_{4})_{21}=[\omega_{c}^{2}(\hat{n}-\frac{1}{2})^{2}+\lambda^{2}\hat{n}+(\frac{\Delta(\vec{p},\vec{g},t)}{4})^{2}]4\omega_{c}(\hat{n}-\frac{1}{2})\kappa\hat{a}^{\dagger},
\end{equation}
\begin{equation}
(\hat{A}_{4})_{22}=-[\omega_{c}^{2}(\hat{n}-\frac{1}{2})^{2}+\lambda^{2}\hat{n}+(\frac{\Delta(\vec{p},\vec{g},t)}{4})^{2}]4\omega_{c}(\hat{n}-\frac{1}{2})(\frac{\Delta(\vec{p},\vec{g},t)}{4}).
\end{equation}
It is easily proved that $[\hat{A}_{3},\hat{A}_{4}]=0$.
Taking into account the initial condition (24) the auxiliary density operator $\hat{\rho}_{2}(t)$ is defined as
\begin{eqnarray}
\hat{\rho}_{2}(t)&&=\exp(\hat{S}t)\exp(\hat{T}t)\hat{\rho}_{s}(0)\\ \nonumber &&
=\exp(-i\hat{H}_{2}t)\exp(-\gamma \hat{A}_{2} t) \exp(-\gamma \eta \hat{A}_{4} t) \\ \nonumber  && \times \hat{\rho}_{1}(t) \exp(-\gamma \eta \hat{A}_{4} t)\exp(-\gamma \hat{A}_{2} t)\exp(i\hat{H}_{2}t),
\end{eqnarray}
where the operator $\hat{\rho}_{1}(t)$ is defined by
\begin{equation}
\hat{\rho}_{1}(t)=|\Psi(t)\rangle\langle\Psi(t)|\otimes |e\rangle\langle e|,
\end{equation}
with
\begin{eqnarray}
|\Psi(t)\rangle &&=\exp(-\gamma (1+\eta)t[\omega_{c}^{2}(\hat{n}+\frac{1}{2})^{2}+\lambda^{2}(\hat{n}+1)+
(\frac{\Delta(\vec{p},\vec{g},t)}{4})^{2}])\\ \nonumber && \times
\exp(-4\gamma \eta \omega_{c}^{2}(\hat{n}+\frac{1}{2})^{2}(\lambda^{2}(\hat{n}+1)+
(\frac{\Delta(\vec{p},\vec{g},t)}{4})^{2}))
 w_{n}(0)\exp(-in\omega_{c}t)|n\rangle.
\end{eqnarray}
From (30) and (33) we have, respectively
\begin{equation}
\exp(-i \hat{H}_{1}t)= \left [\begin{array}{cccc}
 \exp(-i\omega_{c}t(\hat{n}+\frac{1}{2}))& 0 \\
0 & \exp(-i\omega_{c}t(\hat{n}-\frac{1}{2})) \\
\end{array}\right],
\end{equation}
\begin{equation}
\exp(-\gamma \hat{A}_{1}t)= \left [\begin{array}{cccc}
 (\hat{A}_{1})_{11}(\hat{n},t)& 0 \\
0 & (\hat{A}_{1})_{22}(\hat{n},t) \\
\end{array}\right],
\end{equation}
where
\begin{equation}
(\hat{A}_{1})_{11}(\hat{n},t)=\exp(-\gamma t[\omega_{c}^{2}(\hat{n}+\frac{1}{2})^{2}+\lambda^{2}(\hat{n}+1)+(\frac{\Delta(\vec{p},\vec{g},t)}{4})^{2}]),
\end{equation}
\begin{equation}
(\hat{A}_{1})_{22}(\hat{n},t)=\exp(-\gamma t[\omega_{c}^{2}(\hat{n}-\frac{1}{2})^{2}+\lambda^{2}\hat{n}+(\frac{\Delta(\vec{p},\vec{g},t)}{4})^{2}]).
\end{equation}
Also, we can write
\begin{equation}
\exp(-\gamma \hat{A}_{2}t)= \left [\begin{array}{cccc}
\hat{e}_{1}(\hat{n},t)& \hat{e}_{2}(\hat{n},t)\hat{a} \\
 \hat{e}_{3}(\hat{n},t)\hat{a}^{\dagger}& \hat{e}_{4}(\hat{n},t) \\
\end{array}\right],
\end{equation}
\begin{equation}
\exp(-\gamma \eta \hat{A}_{4}t)= \left [\begin{array}{cccc}
\hat{e'}_{1}(\hat{n},t)& \hat{e'}_{2}(\hat{n},t)\hat{a} \\
 \hat{e'}_{3}(\hat{n},t)\hat{a}^{\dagger}& \hat{e'}_{4}(\hat{n},t) \\
\end{array}\right],
\end{equation}
where
\begin{equation}
\hat{e}_{1}(\hat{n},t)=\cosh(\gamma t \sqrt{\hat{c}_{1}(\hat{n},t)} )-\omega_{c}(\frac{\Delta(\vec{p},\vec{g},t)}{2})(\hat{n}+\frac{1}{2} )\frac{\sinh(\gamma t \sqrt{\hat{c}_{1}(\hat{n},t)})}{\sqrt{\hat{c}_{1}(\hat{n},t)}},
\end{equation}
\begin{equation}
\hat{e}_{2}(\hat{n},t)=-2\omega_{c}\lambda(\hat{n}-\frac{1}{2} )\frac{\sinh(\gamma t \sqrt{\hat{c}_{1}(\hat{n}-1,t)})}{\sqrt{\hat{c}_{1}(\hat{n}-1,t)}},
\end{equation}
\begin{equation}
\hat{e}_{3}(\hat{n},t)=-2\omega_{c}\lambda(\hat{n}-\frac{1}{2} )\frac{\sinh(\gamma t \sqrt{\hat{c}_{2}(\hat{n},t)})}{\sqrt{\hat{c}_{2}(\hat{n},t)}},
\end{equation}
\begin{equation}
\hat{e}_{4}(\hat{n},t)=\cosh(\gamma t \sqrt{\hat{c}_{2}(\hat{n},t)} )-\omega_{c}(\frac{\Delta(\vec{p},\vec{g},t)}{2})(\hat{n}-\frac{1}{2} )\frac{\sinh(\gamma t \sqrt{\hat{c}_{2}(\hat{n},t)})}{\sqrt{\hat{c}_{2}(\hat{n},t)}},
\end{equation}
with
\begin{equation}
\hat{c}_{1}(\hat{n},t)=\omega_{c}^{2}(\frac{\Delta(\vec{p},\vec{g},t)}{2})^{2}(\hat{n}+\frac{1}{2})^{2}+\lambda^{2}(\frac{\Delta(\vec{p},\vec{g},t)}{2})^{2}(\hat{n}+1)(\hat{n}+\frac{1}{2})^{2},
\end{equation}
\begin{equation}
\hat{c}_{2}(\hat{n},t)=\omega_{c}^{2}(\frac{\Delta(\vec{p},\vec{g},t)}{2})^{2}(\hat{n}-\frac{1}{2})^{2}+\lambda^{2}(\frac{\Delta(\vec{p},\vec{g},t)}{2})^{2}\hat{n}(\hat{n}-\frac{1}{2})^{2},
\end{equation}
and
\begin{equation}
\hat{e'}_{1}(\hat{n},t)=\cosh(\gamma \eta t \sqrt{\hat{c'}_{1}(\hat{n},t)} )-
2\omega_{c}(\frac{\Delta(\vec{p},\vec{g},t)}{2})(\hat{n}+\frac{1}{2} )\hat{L}_{1}(\hat{n},t)
\frac{\sinh(\gamma \eta t \sqrt{\hat{c'}_{1}(\hat{n},t)})}{\sqrt{\hat{c'}_{1}(\hat{n},t)}},
\end{equation}
\begin{equation}
\hat{e'}_{2}(\hat{n},t)=-2\omega_{c}\lambda(\hat{n}-\frac{1}{2} ) \hat{L}_{2}(\hat{n},t)\frac{\sinh(\gamma \eta t \sqrt{\hat{c'}_{1}(\hat{n}-1,t)})}{\sqrt{\hat{c'}_{1}(\hat{n}-1,t)}},
\end{equation}
\begin{equation}
\hat{e'}_{3}(\hat{n},t)=-2\omega_{c}\lambda(\hat{n}-\frac{1}{2} )\hat{L}_{2}(\hat{n},t)\frac{\sinh(\gamma \eta t \sqrt{\hat{c'}_{2}(\hat{n},t)})}{\sqrt{\hat{c'}_{2}(\hat{n},t)}},
\end{equation}
\begin{equation}
\hat{e'}_{4}(\hat{n},t)=\cosh(\gamma \eta t \sqrt{\hat{c'}_{2}(\hat{n},t)} )-\omega_{c}(\frac{\Delta(\vec{p},\vec{g},t)}{2})(\hat{n}-\frac{1}{2} )\hat{L}_{2}(\hat{n},t)\frac{\sinh(\gamma \eta t \sqrt{\hat{c'}_{2}(\hat{n},t)})}{\sqrt{\hat{c'}_{2}(\hat{n},t)}},
\end{equation}
with
\begin{equation}
\hat{c'}_{1}(\hat{n},t)=\omega_{c}^{2}(\frac{\Delta(\vec{p},\vec{g},t)}{2})^{2}(\hat{n}+\frac{1}{2})^{2}
\hat{L}_{1}^{2}(\hat{n},t)+\hat{L}_{1}(\hat{n},t)\hat{L}_{2}(\hat{n},t)(\hat{n}+1)(\hat{n}+\frac{1}{2})^{2},
\end{equation}
\begin{equation}
\hat{c'}_{2}(\hat{n},t)=\omega_{c}^{2}(\frac{\Delta(\vec{p},\vec{g},t)}{2})^{2}(\hat{n}
-\frac{1}{2})^{2}\hat{L}_{2}^{2}(\hat{n},t)+\hat{L}_{2}(\hat{n},t)\hat{L}_{1}(\hat{n},t)\hat{n}(\hat{n}-\frac{1}{2})^{2},
\end{equation}
where
\begin{equation}
\hat{L}_{1}(\hat{n},t)=\omega_{c}^{2}(\hat{n}+\frac{1}{2})^{2}+\lambda^{2}(\hat{n}+1)+(\frac{\Delta(\vec{p},\vec{g},t)}{2})^{2},
\end{equation}
\begin{equation}
\hat{L}_{2}(\hat{n},t)=\omega_{c}^{2}(\hat{n}-\frac{1}{2})^{2}+\lambda^{2}\hat{n}+(\frac{\Delta(\vec{p},\vec{g},t)}{2})^{2}.
\end{equation}
Similarly, the operator $\exp(-i \hat{H}_{2}t)$ in the two-dimensional atomic basis can be stated as
\begin{equation}
\exp(-i \hat{H}_{2}t)= \left [\begin{array}{cccc}
\hat{d}_{1}(\hat{n},t)&\hat{d}_{2}(\hat{n},t)\hat{a} \\
 \hat{d}_{3}(\hat{n},t)\hat{a}^{\dagger}& \hat{d}_{4}(\hat{n},t) \\
\end{array}\right],
\end{equation}
where
\begin{equation}
\hat{d}_{1}(\hat{n},t)=\cos(t((\frac{\Delta(\vec{p},\vec{g},t)}{4})^{2} +\lambda^{2}(\hat{n}+1)))-(\frac{\Delta(\vec{p},\vec{g},t)}{4})\frac{\sin(t((\frac{\Delta(\vec{p},\vec{g},t)}{4})^{2} +\lambda^{2}(\hat{n}+1)))}{\sqrt{(\frac{\Delta(\vec{p},\vec{g},t)}{4})^{2} +\lambda^{2}(\hat{n}+1))}},
\end{equation}
\begin{equation}
\hat{d}_{2}(\hat{n},t)=-i\lambda \frac{\sin(t((\frac{\Delta(\vec{p},\vec{g},t)}{4})^{2} +\lambda^{2}(\hat{n}+1)))}{\sqrt{(\frac{\Delta(\vec{p},\vec{g},t)}{4})^{2} +\lambda^{2}(\hat{n}+1)}},
\end{equation}
\begin{equation}
\hat{d}_{3}(\hat{n},t)=-i\lambda \frac{\sin(t((\frac{\Delta(\vec{p},\vec{g},t)}{4})^{2} +\lambda^{2}\hat{n}))}{\sqrt{(\frac{\Delta(\vec{p},\vec{g},t)}{4})^{2} +\lambda^{2}\hat{n}}},
\end{equation}
\begin{equation}
\hat{d}_{4}(\hat{n},t)=\cos(t((\frac{\Delta(\vec{p},\vec{g},t)}{4})^{2} +\lambda^{2}\hat{n}))-(\frac{\Delta(\vec{p},\vec{g},t)}{4})\frac{\sin(t((\frac{\Delta(\vec{p},\vec{g},t)}{4})^{2} +\lambda^{2}\hat{n}))}{\sqrt{(\frac{\Delta(\vec{p},\vec{g},t)}{4})^{2} +\lambda^{2}\hat{n})}}.
\end{equation}
Then, from (51) and (67), it follows that
\begin{equation}
\exp(-i \hat{H}_{2}t)\exp(-\gamma \hat{A}_{2}t)= \left [\begin{array}{cccc}
\hat{f}_{1}(\hat{n},t)&\hat{f}_{2}(\hat{n},t)\hat{a} \\
 \hat{f}_{3}(\hat{n},t)\hat{a}^{\dagger}& \hat{f}_{4}(\hat{n},t) \\
\end{array}\right],
\end{equation}
where
\begin{equation}
\hat{f}_{1}(\hat{n},t)=\hat{e}_{1}(\hat{n},t)\hat{d}_{1}(\hat{n},t)+\hat{e}_{2}(\hat{n},t)\hat{d}_{2}(\hat{n},t),
\end{equation}
\begin{equation}
\hat{f}_{2}(\hat{n},t)=\hat{e}_{2}(\hat{n},t)\hat{d}_{1}(\hat{n},t)+\hat{e}_{1}(\hat{n},t)\hat{d}_{2}(\hat{n},t),
\end{equation}
\begin{equation}
\hat{f}_{3}(\hat{n},t)=\hat{e}_{3}(\hat{n},t)\hat{d}_{4}(\hat{n},t)+\hat{e}_{4}(\hat{n},t)\hat{d}_{3}(\hat{n},t),
\end{equation}
\begin{equation}
\hat{f}_{4}(\hat{n},t)=\hat{e}_{4}(\hat{n},t)\hat{d}_{4}(\hat{n},t)+\hat{e}_{3}(\hat{n},t)\hat{d}_{3}(\hat{n},t).
\end{equation}
Also, from (52) and (60), we have
\begin{equation}
\exp(-i \hat{H}_{2}t)\exp(-\gamma \hat{A}_{2}t)\exp(-\gamma \eta \hat{A}_{4}t)= \left [\begin{array}{cccc}
\hat{J}_{1}(\hat{n},t)&\hat{J}_{2}(\hat{n},t)\hat{a} \\
 \hat{J}_{3}(\hat{n},t)\hat{a}^{\dagger}& \hat{J}_{4}(\hat{n},t) \\
\end{array}\right],
\end{equation}
where
\begin{equation}
\hat{J}_{1}(\hat{n},t)=\hat{f}_{1}(\hat{n},t)\hat{e'}_{1}(\hat{n},t)+\hat{f}_{2}(\hat{n},t)\hat{e'}_{2}(\hat{n},t),
\end{equation}
\begin{equation}
\hat{J}_{2}(\hat{n},t)=\hat{f}_{2}(\hat{n},t)\hat{e'}_{1}(\hat{n},t)+\hat{f}_{1}(\hat{n},t)\hat{e'}_{2}(\hat{n},t),
\end{equation}
\begin{equation}
\hat{J}_{3}(\hat{n},t)=\hat{f}_{3}(\hat{n},t)\hat{e'}_{4}(\hat{n},t)+\hat{f}_{4}(\hat{n},t)\hat{e'}_{3}(\hat{n},t),
\end{equation}
\begin{equation}
\hat{J}_{4}(\hat{n},t)=\hat{f}_{4}(\hat{n},t)\hat{e'}_{4}(\hat{n},t)+\hat{f}_{3}(\hat{n},t)\hat{e'}_{3}(\hat{n},t).
\end{equation}
Substituting (45) and (77) into (44), an explicit expression for the operator $\hat{\rho}_{2}(t)$ can be obtained
as follows
\begin{equation}
(\hat{\rho}_{2}(t))_{i,j}=|\Psi_{i}(t)\rangle\langle\Psi_{j}(t)|,(i,j=1,2),
\end{equation}
with
\begin{equation}
|\Psi_{1}(t)\rangle=\hat{J}_{1}(\hat{n},t)|\Psi(t)\rangle, |\Psi_{2}(t)\rangle=\hat{J}_{3}(\hat{n},t)|\Psi(t)\rangle,
\end{equation}
where $|\Psi(t)\rangle$ is given by Eq.(46).
Now, we obtain the action of the operator $\exp(\hat{R}t)=\exp(\hat{R_{1}}t)\exp(\hat{R_{2}}t)\exp(\hat{R_{3}}t)$
 on the operator $\hat{\rho}_{2}(t)$
\begin{equation}
\hat{\rho}_{3}(t)=\sum_{i=0}^{\infty}\frac{(- \gamma \eta t)^{i} }{i!}\hat{H}^{i}\hat{\rho}_{2}(t)\hat{H}^{2i},
\end{equation}
\begin{equation}
\hat{\rho}_{4}(t)=\sum_{j=0}^{\infty}\frac{(-3 \gamma \eta t)^{j} }{j!}\hat{H}^{2j}\hat{\rho}_{3}(t)\hat{H}^{j},
\end{equation}
\begin{equation}
\hat{\rho}_{s}(t)=\sum_{k=0}^{\infty}\frac{(2 \gamma t)^{k} }{k!}\hat{H}^{k}\hat{\rho}_{4}(t)\hat{H}^{k},
\end{equation}
where
\begin{equation}
\hat{H}^{k}=\sum_{l=0}^{k}\frac{k! }{l!(k-l)!}\hat{H}_{1}^{k-l}\hat{H}_{2}^{l},
\end{equation}
which can be explicitly expressed as follows
\begin{equation}
\hat{H}^{k}= \left [\begin{array}{cccc}
\hat{g}_{+}^{k} (\hat{n},t)&\kappa^{*}(t) \frac{\hat{u}_{-}^{k} (\hat{n},t)}{\sqrt{(\frac{\Delta(\vec{p},\vec{g},t)}{4})^{2} +\lambda^{2}(\hat{n}+1))}} \hat{a} \\
\kappa(t)\frac{\hat{v}_{-}^{k} (\hat{n},t)}{\sqrt{(\frac{\Delta(\vec{p},\vec{g},t)}{4})^{2} +\lambda^{2}(\hat{n}+1))}} \hat{a}^{\dagger}&\hat{g}_{-}^{k} (\hat{n},t)  \\
\end{array}\right],
\end{equation}
where
\begin{equation}
\hat{g}_{+}^{k} (\hat{n},t)=\hat{u}_{+}^{k} (\hat{n},t)+\frac{\Delta(\vec{p},\vec{g},t)}{4}\hat{u}_{-}^{k} (\hat{n},t),
\end{equation}
\begin{equation}
\hat{g}_{-}^{k} (\hat{n},t)=\hat{v}_{+}^{k} (\hat{n},t)-\frac{\Delta(\vec{p},\vec{g},t)}{4}\hat{v}_{-}^{k} (\hat{n},t),
\end{equation}
\begin{equation}
\hat{u}_{\pm}^{k} (\hat{n},t)=\frac{1}{2}(\hat{r}_{+}^{k} (\hat{n},t) \pm \hat{r}_{-}^{k} (\hat{n},t)),\hat{v}_{\pm}^{k} (\hat{n},t)=\frac{1}{2}(\hat{s}_{+}^{k} (\hat{n},t) \pm \hat{s}_{-}^{k} (\hat{n},t)),
\end{equation}
with
\begin{equation}
\hat{r}_{\pm} (\hat{n},t)=\omega_{c}(\hat{n}+\frac{1}{2} )\pm\sqrt{(\frac{\Delta(\vec{p},\vec{g},t)}{4})^{2} +\lambda^{2}(\hat{n}+1))},
\end{equation}
\begin{equation}
\hat{s}_{\pm} (\hat{n},t)=\omega_{c}(\hat{n}-\frac{1}{2} )\pm\sqrt{(\frac{\Delta(\vec{p},\vec{g},t)}{4})^{2} +\lambda^{2}\hat{n}}.
\end{equation}
Finally, by substituting (85) and (87) into (86) we obtain the exact solution of
the non-Markovian master equation (12) for the phase
damped JCM in the presence of a classical homogeneous gravity field
\begin{equation}
\hat{\rho}_{s}(t)= \left [\begin{array}{cccc}
 (\hat{\rho}_{s})_{11}(t) & (\hat{\rho}_{s})_{12}(t)\\
 (\hat{\rho}_{s})_{21}(t) &(\hat{\rho}_{s})_{22}(t)  \\
\end{array}\right],
\end{equation}
where
\begin{eqnarray}
(\hat{\rho}_{s})_{11}(t) =&&\sum_{k=0}^{\infty}\frac{(2 \gamma t)^{k} }{k!}
(\hat{g}_{+}^{k} (\hat{n},t)(\hat{\rho}_{4})_{11}(t)\hat{g}_{+}^{k} (\hat{n},t)
\\ \nonumber+&&\hat{a}\hat{v}_{-}^{'k} (\hat{n},t)(\hat{\rho}_{4})_{21}(t)\hat{g}_{+}^{k} (\hat{n},t)
\\ \nonumber+&&\hat{g}_{+}^{k} (\hat{n},t)(\hat{\rho}_{4})_{12}(t)\hat{v}_{-}^{'k} (\hat{n},t)\hat{a}^{\dagger}
\\ \nonumber+&&\hat{a}\hat{v}_{-}^{'k} (\hat{n},t)(\hat{\rho}_{4})_{22}(t)\hat{v}_{-}^{'k} (\hat{n},t)\hat{a}^{\dagger})|\phi(\vec{p})|^{2},
\end{eqnarray}
\begin{eqnarray}
(\hat{\rho}_{s})_{22}(t) =&&\sum_{k=0}^{\infty}\frac{(2 \gamma t)^{k} }{k!}
(\hat{v}_{-}^{'k} (\hat{n},t)\hat{a}^{\dagger}(\hat{\rho}_{4})_{11}(t)\hat{a}\hat{v}_{-}^{'k} (\hat{n},t)
\\ \nonumber+&&\hat{g}_{-}^{k} (\hat{n},t)(\hat{\rho}_{4})_{21}(t)\hat{a}\hat{v}_{+}^{'k} (\hat{n},t)
\\ \nonumber+&&\hat{v}_{-}^{'k} (\hat{n},t)\hat{a}^{\dagger}(\hat{\rho}_{4})_{12}(t)\hat{g}_{-}^{k} (\hat{n},t)
\\ \nonumber+&&\hat{g}_{-}^{k} (\hat{n},t)(\hat{\rho}_{4})_{22}(t)\hat{g}_{-}^{k} (\hat{n},t))|\phi(\vec{p})|^{2},
\end{eqnarray}
\begin{eqnarray}
(\hat{\rho}_{s})_{12}(t)=&&(\hat{\rho}_{s})_{21}(t)^{\dagger}=\sum_{k=0}^{\infty}\frac{(2 \gamma t)^{k} }{k!}
(\hat{v}_{-}^{'k} (\hat{n},t)\hat{a}^{\dagger}(\hat{\rho}_{4})_{11}(t)\hat{g}_{+}^{k} (\hat{n},t)
\\ \nonumber+&&\hat{g}_{-}^{k} (\hat{n},t)(\hat{\rho}_{4})_{21}(t) \hat{a}\hat{g}_{+}^{k} (\hat{n},t)
\\ \nonumber+&&\hat{v}_{-}^{'k} (\hat{n},t)\hat{a}^{\dagger}(\hat{\rho}_{4})_{12}(t)(t)
\hat{v}_{-}^{'k} (\hat{n},t)\hat{a}^{\dagger}
\\ \nonumber+&&\hat{g}_{-}^{k} (\hat{n},t)(\hat{\rho}_{4})_{22}(t)\hat{v}_{-}^{'k} (\hat{n},t)
\hat{a}^{\dagger})|\phi(\vec{p})|^{2},
\end{eqnarray}
with
\begin{equation}
\hat{v}_{-}^{'k} (\hat{n},t)=\frac{\lambda}{\sqrt{(\frac{\Delta(\vec{p},\vec{g},t)}{4} )^{2}
+ \lambda^{2}\hat{n} }}\hat{v}_{-}^{k} (\hat{n},t)
\end{equation}
where
\begin{eqnarray}
(\hat{\rho}_{4})_{11}(t) =&&\sum_{j=0}^{\infty}\frac{(-3 \gamma \eta t)^{j} }{j!}
(\hat{g}_{+}^{2j} (\hat{n},t)(\hat{\rho}_{3})_{11}(t)\hat{g}_{+}^{j} (\hat{n},t)
\\ \nonumber+&&\hat{a}\hat{v}_{-}^{'2j} (\hat{n},t)(\hat{\rho}_{3})_{21}(t)\hat{g}_{+}^{j} (\hat{n},t)
\\ \nonumber+&&\hat{g}_{+}^{2j} (\hat{n},t)(\hat{\rho}_{3})_{12}(t)\hat{v}_{-}^{'j} (\hat{n},t)\hat{a}^{\dagger}
\\ \nonumber+&&\hat{a}\hat{v}_{-}^{'2j} (\hat{n},t)(\hat{\rho}_{3})_{22}(t)\hat{v}_{-}^{'j} (\hat{n},t)\hat{a}^{\dagger}),
\end{eqnarray}
\begin{eqnarray}
(\hat{\rho}_{4})_{22}(t) =&&\sum_{j=0}^{\infty}\frac{(-3 \gamma \eta t)^{j} }{j!}
(\hat{v}_{-}^{'2j} (\hat{n},t)\hat{a}^{\dagger}(\hat{\rho}_{3})_{11}(t)\hat{a}\hat{v}_{-}^{'j} (\hat{n},t)
\\ \nonumber+&&\hat{g}_{-}^{2j} (\hat{n},t)(\hat{\rho}_{3})_{21}(t)\hat{a}\hat{v}_{+}^{'j} (\hat{n},t)
\\ \nonumber+&&\hat{v}_{-}^{'2j} (\hat{n},t)\hat{a}^{\dagger}(\hat{\rho}_{3})_{12}(t)\hat{g}_{-}^{j} (\hat{n},t)
\\ \nonumber+&&\hat{g}_{-}^{2j} (\hat{n},t)(\hat{\rho}_{3})_{22}(t)\hat{g}_{-}^{j} (\hat{n},t)),
\end{eqnarray}
\begin{eqnarray}
(\hat{\rho}_{4})_{12}(t)=&&(\hat{\rho}_{4})_{21}(t)^{\dagger}\\ \nonumber=&&\sum_{j=0}^{\infty}\frac{(-3 \gamma \eta t)^{j} }{j!}
(\hat{v}_{-}^{'2j} (\hat{n},t)\hat{a}^{\dagger}(\hat{\rho}_{3})_{11}(t)\hat{g}_{+}^{j} (\hat{n},t)
\\ \nonumber+&&\hat{g}_{-}^{2j} (\hat{n},t)(\hat{\rho}_{3})_{21}(t) \hat{a}\hat{g}_{+}^{j} (\hat{n},t)
\\ \nonumber+&&\hat{v}_{-}^{'2j} (\hat{n},t)\hat{a}^{\dagger}(\hat{\rho}_{3})_{12}(t)(t)
\hat{v}_{-}^{'j} (\hat{n},t)\hat{a}^{\dagger}
\\ \nonumber+&&\hat{g}_{-}^{2j} (\hat{n},t)(\hat{\rho}_{3})_{22}(t)\hat{v}_{-}^{'j} (\hat{n},t)
\hat{a}^{\dagger}),
\end{eqnarray}
with
\begin{eqnarray}
(\hat{\rho}_{3})_{11}(t) =&&\sum_{i=0}^{\infty}\frac{(- \gamma \eta t)^{i} }{i!}
(\hat{g}_{+}^{i} (\hat{n},t)(\hat{\rho}_{2})_{11}(t)\hat{g}_{+}^{2i} (\hat{n},t)
\\ \nonumber+&&\hat{a}\hat{v}_{-}^{'i} (\hat{n},t)(\hat{\rho}_{2})_{21}(t)\hat{g}_{+}^{2i} (\hat{n},t)
\\ \nonumber+&&\hat{g}_{+}^{i} (\hat{n},t)(\hat{\rho}_{2})_{12}(t)\hat{v}_{-}^{'2i} (\hat{n},t)\hat{a}^{\dagger}
\\ \nonumber+&&\hat{a}\hat{v}_{-}^{'i} (\hat{n},t)(\hat{\rho}_{2})_{22}(t)\hat{v}_{-}^{'2i} (\hat{n},t)\hat{a}^{\dagger}),
\end{eqnarray}
\begin{eqnarray}
(\hat{\rho}_{3})_{22}(t) =&&\sum_{i=0}^{\infty}\frac{(- \gamma \eta t)^{i} }{i!}
(\hat{v}_{-}^{'i} (\hat{n},t)\hat{a}^{\dagger}(\hat{\rho}_{2})_{11}(t)\hat{a}\hat{v}_{-}^{'2i} (\hat{n},t)
\\ \nonumber+&&\hat{g}_{-}^{i} (\hat{n},t)(\hat{\rho}_{2})_{21}(t)\hat{a}\hat{v}_{+}^{'2i} (\hat{n},t)
\\ \nonumber+&&\hat{v}_{-}^{'i} (\hat{n},t)\hat{a}^{\dagger}(\hat{\rho}_{2})_{12}(t)\hat{g}_{-}^{2i} (\hat{n},t)
\\ \nonumber+&&\hat{g}_{-}^{i} (\hat{n},t)(\hat{\rho}_{2})_{22}(t)\hat{g}_{-}^{2i} (\hat{n},t)),
\end{eqnarray}
\begin{eqnarray}
(\hat{\rho}_{3})_{12}(t)=&&(\hat{\rho}_{4})_{21}(t)^{\dagger}\\ \nonumber=&&\sum_{i=0}^{\infty}\frac{(- \gamma \eta t)^{i} }{i!}
(\hat{v}_{-}^{'i} (\hat{n},t)\hat{a}^{\dagger}(\hat{\rho}_{2})_{11}(t)\hat{g}_{+}^{2i} (\hat{n},t)
\\ \nonumber+&&\hat{g}_{-}^{i} (\hat{n},t)(\hat{\rho}_{2})_{21}(t) \hat{a}\hat{g}_{+}^{2i} (\hat{n},t)
\\ \nonumber+&&\hat{v}_{-}^{'i} (\hat{n},t)\hat{a}^{\dagger}(\hat{\rho}_{2})_{12}(t)(t)
\hat{v}_{-}^{'2i} (\hat{n},t)\hat{a}^{\dagger}
\\ \nonumber+&&\hat{g}_{-}^{i} (\hat{n},t)(\hat{\rho}_{2})_{22}(t)\hat{v}_{-}^{'2i} (\hat{n},t)
\hat{a}^{\dagger}),
\end{eqnarray}
where we have defined $(\hat{\rho}_{2}(t))_{i,j},(i,j=1,2)$ in (82).\\
\hspace*{00.5 cm}Making use of the solution given by (94), one can evaluate the mean values of operators of interest.
 In the next section, it will be used to investigate various dynamical properties of the non-Markovian phase damped
 JCM in the presence of a homogeneous
gravitational field.
\section{Dynamical Properties}
In this section, the non-Markovian effects on the quantum statistical properties of the atom and the
 quantized radiation field in the presence of both the phase damping and the gravitational
field will be studied.  \\
 \\
 \\
 \\
 \\
{\bf  4a. Atomic Population Inversion} \\
\\
The atomic population inversion is expressed by the expression
\begin{equation}
W(t)=\langle\hat{\sigma}_{3}(t)\rangle=Tr_{atom}(\hat{\rho}_{atom}(t)\hat{\sigma}_{3}(t)),
\end{equation}
where
\begin{equation}
\hat{\rho}_{atom}(t)=Tr_{field}(\hat{\rho}_{s}(t).
\end{equation}
It can be rewritten (105) as follows
\begin{equation}
W(t)=\int d^{3}p\sum_{i=e,g}\langle i|\hat{\rho}_{atom}(t)\hat{\sigma}_{3}(t)|i \rangle=\int d^{3}p\sum_{n=0}^{\infty}(\langle n|\otimes(\langle e|\hat{\rho}_{s}(t)|e \rangle- \langle g|\hat{\rho}_{s}(t)|g \rangle    )\otimes|n \rangle ).
\end{equation}
Therefore, by using (94) and (107) we obtain
\begin{equation}
W(t)=\int d^{3}p (\sum_{k=0}^{\infty}\sum_{n=0}^{\infty} \frac{(2 \gamma t)^{k} }{k!}(\langle n | (\hat{\rho}_{s})_{11}(t)|n \rangle -\langle n | (\hat{\rho}_{s})_{22}(t)|n \rangle )),
\end{equation}
where $(\hat{\rho}_{2}(t))_{i,j},(i,j=1,2)$ is given by (82).\\
 \hspace*{00.5 cm} In Fig.1, the atomic population inversion as a
 function of the scaled time $\lambda t$ for different values of the
 parameters $\vec{q}.\vec{g}$, $\gamma$ and $\eta$ are plotted. In this figure and all the
 subsequent figures we set
$q=10^{7}m^{-1}$, $M=10^{-26}Kg$, $g=9.8\frac{m}{s^{2}}$,
$\omega_{rec}=\frac{\hbar q^{2}}{2M}=.5\times10^{6}\frac{rad}{s}$,
$\lambda=1\times 10^{6}\frac{rad}{s}$, $ \alpha=2$,
$\Delta=1.8\times 10^{6}\frac{rad}{s}$,
$\phi(\vec{p})=\frac{1}{\sqrt{2\pi
\sigma_{0}}}\exp(\frac{-p^{2}}{\sigma_{0}^{2}})$ with
$\sigma_{0}=1$ [25-30]. In Fig.1a, three parameters: $\vec{q}.\vec{g}=0$, $\gamma=0$ and $\eta=0$ are considered.
 When $\vec{q}.\vec{g}=0$, the
momentum transfer from the laser beam to the atom is only
slightly altered by the gravitational acceleration because the
latter is very small or nearly perpendicular to the laser beam. When $\gamma=0$, there is no the phase damping and
$\eta=0$ means that we consider Markovian approach. As it is seen from Fig.1a for the atomic
population inversion the Rabi-like oscillations can be
identified.
In Figs.1b and 1c, small gravitational influence $\vec{q}.\vec{g}=0.1\times 10^{7}$ with $\eta=0$
for $\gamma=0$ and $\gamma=7\times10^{-5}\frac{rad}{s}$ is considered, respectively.
By considering both small gravitational influence and small non-Markovian effect in the presence of the phase damping
the Rabi oscillations of
the atomic population inversion disappear (see Fig.1d).
  By comparing Figs.1c,1d and 1f, the influence of the non-Markovian on the time evolution of the atomic
population inversion when there are both the phase damping and the gravitational field can be seen.
\\
  \\
{\bf  4b. Photon Counting Statistics} \\
\\
  Now, the influence of gravity on the sub-Poissonian statistics of the radiation
 field will be investigated. For this purpose, we calculate the Mandel parameter defined by [46]
\begin{equation}
Q(t)=\frac{(\langle n(t)^{2}\rangle-\langle n(t)\rangle
^{2})}{\langle n(t)\rangle}-1.
\end{equation}
For $Q<0$ $(Q>0)$, the statistics is sub-Poissonian
(super-Poissonian); $Q=0$ stands for Poissonian statistics. Since
$\langle n(t)\rangle=\sum_{n=0}^{\infty}n P(n,t)$ and $\langle
n(t)^{2}\rangle=\sum_{n=0}^{\infty}n^{2} P(n,t)$ we have
\begin{equation}
Q(t)=(\{[\sum_{n=0}^{\infty} n^{2}P(n,t)
]-[\sum_{n=0}^{\infty} n P(n,t) ]^{2}    \}
[\sum_{n=0}^{\infty} n P(n,t) ]^{-1})-1,
\end{equation}
where the probability of finding $n$ photons in the radiation field is found to be
\begin{equation}
P(n,t)=\langle n|\hat{\rho}_{field}(t)|n \rangle=\langle n|Tr_{atom} \hat{\rho}_{s}(t)|n \rangle,
\end{equation}
and by using (94) we have
\begin{equation}
P(n,t)=\int d^{3}p\sum_{k=0}^{\infty}\sum_{n=0}^{\infty} \frac{(2 \gamma t)^{k} }{k!}(\langle n | (\hat{\rho}_{s})_{11}(t)|n \rangle +\langle n | (\hat{\rho}_{s})_{22}(t)|n \rangle ).
\end{equation}
Therefore, by using (110) and (112) we obtain\\
\\
\\
\\
\begin{eqnarray}
Q(t)&&=(\{[\sum_{n=0}^{\infty} n^{2}(\int d^{3}p\sum_{k=0}^{\infty}\sum_{n=0}^{\infty} \frac{(2 \gamma t)^{k} }{k!}
\\ \nonumber && \times(\langle n | (\hat{\rho}_{s})_{11}(t)|n \rangle +\langle n | (\hat{\rho}_{s})_{22}(t)|n \rangle ))
]\\ \nonumber &&-[\sum_{n=0}^{\infty} n (\int d^{3}p\sum_{k=0}^{\infty}\sum_{n=0}^{\infty} \frac{(2 \gamma t)^{k} }{k!}(\langle n | (\hat{\rho}_{s})_{11}(t)|n \rangle +\langle n |(\hat{\rho}_{s})_{22}(t)|n \rangle )) ]^{2}    \}
\\ \nonumber &&\times[\sum_{n=0}^{\infty} n (\int d^{3}p\sum_{k=0}^{\infty}\sum_{n=0}^{\infty} \frac{(2 \gamma t)^{k} }{k!}(\langle n |(\hat{\rho}_{s})_{11}(t)|n \rangle +\langle n | (\hat{\rho}_{s})_{22}(t)|n \rangle )) ]^{-1})-1.
\end{eqnarray}
\hspace*{00.5 cm}The numerical results for three values of the parameter $\vec{q}.\vec{g}$, $\gamma$ and $\eta$
are shown in Fig.2. As
it is seen, the cavity-field exhibits alternately sub-Poissonian and
super-Poissonian statistics when every three influences are negligible. For small gravitational influence and
small non-Markovian effect in the presence of the phase damping or with increasing
$\eta$ the sub-Poissonian characteristic is suppressed
and the cavity-field exhibits super-Poissonian statistics.
 After some time, the Mandel parameter $Q$ is stabilized at an asymptotic
zero value; the larger the parameter $\eta$ is more rapidly $Q(t)$ reaches the asymptotic value zero.\\
\section{Summary and conclusions}
In this paper, the non-Markovian dissipative dynamics of the phase damped Jaynes-Cummings
           model in the presence of a classical
           homogeneous gravitational field have been analyzed. The model consists of a moving two-level atom simultaneously
exposed to the gravitational field and a single-mode traveling radiation field in the presence of a
non-Markovian phase damping mechanism.
 First, the non-Markovian master equation for the reduced density operator of the system in terms of a
 Hamiltonian describing the atom-field interaction in the presence of a homogeneous gravitational field has been presented.
 Then, the super-operator technique is generalized and an exact solution
 of the non-Markovian master equation is obtained.
 Assuming that initially the radiation
field is prepared in a Glauber coherent state and the two-level atom is
in the excited state, the non-Markovian effects
 on the temporal evolution of collapses and revivals of the atomic
population inversion and photon counting statistics of
the radiation field in the presence of both the phase damping and a homogeneous gravitational field have been investigated.
The results are summarized as follows: with increase of the non-Markovian parameter $\eta$
for small values of the damping parameter $\gamma$ and gravity-dependent parameter $\vec{q}.\vec{g}$,
 1) the Rabi-like oscillations
in the atomic population inversion disappear and 2) the
sub-Poissonian behaviour of the cavity-field is suppressed and it
exhibits super-Poissonian statistics and after some time, the Mandel parameter $Q(t)$ is stabilized at an asymptotic
zero value; the larger the parameter $\eta$ increases $Q(t)$.\\
\\
{\bf  Acknowledgements} \\
The author wishes to thank The Office of Research
of the Islamic Azad University - Shahreza Branch for their
support.

\vspace{20cm}

{\bf FIGURE CAPTIONS:}

{\bf FIG. 1 } Time evolution of the atomic population inversion
versus the scaled time $\lambda t$. Here we have set
$q=10^{7}m^{-1}$,\\
$M=10^{-26}kg$,$g=9.8\frac{m}{s^{2}}$,$\omega_{rec}=.5\times10^{6}\frac{rad}{s}$,\\$\lambda=1\times
10^{6}\frac{rad}{s}$,
 $ \varphi=0$, $\alpha=2$,
$\Delta=1.8\times 10^{6}\frac{rad}{s}$,\\

 {\bf a)}For $\vec{q}.\vec{g}=0$, $\gamma=0$, $\eta=0$.

{\bf b)}For $\vec{q}.\vec{g}=0.1 \times 10^{7}$, $\gamma=0$, $\eta=0$.

{\bf c)}For $\vec{q}.\vec{g}=0.1 \times 10^{7}$, $\gamma=7\times10^{-5}\frac{rad}{s}$, $\eta=0$.

{\bf d)}For $\vec{q}.\vec{g}=0.1 \times 10^{7}$, $\gamma=7\times10^{-5}\frac{rad}{s}$, $\eta=5\times 10^{-5 }\frac{rad}{s}$.

{\bf f)}For $\vec{q}.\vec{g}=0.1 \times 10^{7}$, $\gamma=7\times10^{-5}\frac{rad}{s}$, $\eta=5\times 10^{-3 }\frac{rad}{s}$.\\

{\bf FIG. 2 } Time evolution of the atomic dipole squeezing versus
the scaled time $\lambda t$ with the same corresponding data
 used in Fig.1;\\

{\bf a)}For $\vec{q}.\vec{g}=0$, $\gamma=0$, $\eta=0$.

{\bf b)}For $\vec{q}.\vec{g}=0.1 \times 10^{7}$, $\gamma=0$, $\eta=0$.

{\bf c)}For $\vec{q}.\vec{g}=0.1 \times 10^{7}$, $\gamma=7\times10^{-5}\frac{rad}{s}$, $\eta=0$.

{\bf d)}For $\vec{q}.\vec{g}=0.1 \times 10^{7}$, $\gamma=7\times10^{-5}\frac{rad}{s}$, $\eta=5\times 10^{-5 }\frac{rad}{s}$.

{\bf f)}For $\vec{q}.\vec{g}=0.1 \times 10^{7}$, $\gamma=7\times10^{-5}\frac{rad}{s}$, $\eta=5\times 10^{-3 }\frac{rad}{s}$.\\

\end{document}